\documentclass{article}
\usepackage[utf8]{inputenc}
\usepackage{amsmath, amssymb, amsfonts}
\usepackage{graphicx}
\usepackage{hyperref}
\usepackage{float}
\usepackage{caption}
\usepackage{subcaption}
\usepackage{siunitx}
\usepackage{booktabs}
\usepackage{array}
\usepackage{geometry}
\usepackage{multirow}
\usepackage{setspace}
\begin{document}

\begin{center}
{\Large \textbf{Magnetoelectric Coupling in Nickel-Cobalt Ferrite and Lanthanum Ferrite Heterostructure Composites: Experimental Evidence and Simulation-Driven Insights}}
\vspace{0.3cm}

Manjeet Seth\\
Department of Pure and Applied Physics, Guru Ghasidas Vishwavidyalaya (CU), Koni Bilaspur-495009, C.G., India.\\
E-mail address: manjeetseth28@gmail.com
\end{center}

\vspace{-0.5cm}

\section*{Abstract}
A material that reveals two or more ferroelectric properties at the same time is called multiferroic materials. The most commonly multiferroic materials shows ferroelectricity and ferromagnetism property within a single phase. Accordingly such materials can possess both spontaneous magnetization and electric polarization and which can be individually controlled through external electric or magnetic fields. The unique coexistence of ferroelectric properties opens up possibilities for innovative applications, including memory devices, sensors, and actuators that are responsive to both electrical and magnetic stimuli. Multiferroics gives a valuable foundation for generating cutting-edge multifunctional devices, exhibiting their versatility across broad area technological fields. Applications include sensors, transducers, spintronics, terahertz emitters, miniature antennas, energy harvesting, multiple-state memory storage, electric-field controlled ferromagnetic resonance devices, and nanoscale electronics. Researchers are continually work to discovering new materials exploring the fundamental mechanisms involved, and optimizing their performance for various application. Multiferroic materials has a promising area for innovation and exploration the field of advance technologies this type of materials contributing not only to device development but also enhancing the understanding of the interactions between ferroelectric and magnetic orders. Additionally, magnetoelectric composites is a type of multifunctional material showing strong coupling between magnetic and electric properties for further technological applications. Magnetoelectric (ME) composites are synthesis by combining separate magnetic and electric materials to produce unique functionalities through their interactions. This coupling enables the control of magnetic properties by an electric field and vice versa. The effect arises due to field or strain mediated interactions between the magnetic and electric phases in the composite. Efficient coupling in ME composites that hold great promise for innovative device applications. Many studies have focused on lead based ME composites, but toxic property of the lead concerns about led to the development of lead-free alternatives such as nickel cobalt ferrite (NCFO) and lanthanum ferrite (LaFeO$_3$) based composites.

In the research work an attempt has been made to design and fabricate novel magnetoelectric composites by integrating two distinct functional phases nickel-cobalt ferrite (NCFO), which serves as the magnetostrictive component, and Lanthanum ferrite (LFO), which plays the role of the ferroelectric phase. For the synthesis of composites adopting the conventional solid-state reaction route this method is widely used for preparing ceramic materials because its simplicity and reliability is attracted to researchers for synthesis homogeneous phase formation. The main focus of this investigation is comprehensive examination of the structural characteristics, surface morphology, ferroelectric, magnetoelectric, mössbauer and magnetic behavior of the prepared NCFO-LFO composite materials. The broader aim of characteristics exploring their potential suitability for multifunctional applications in the field of magnetoelectric coupling.

\section{Introduction}
A material that reveals two or more ferroelectric properties at the same time, that is called multiferroic materials. The most commonly multiferroic materials shows ferroelectricity and ferromagnetism property within a single phase~\cite{Wermuth2021}. Accordingly such materials can possess both spontaneous magnetization and electric polarization and which can be individually controlled through external electric or magnetic fields. The unique coexistence of ferroelectric properties opens up possibilities for innovative applications, including memory devices, sensors, and actuators that are responsive to both electrical and magnetic stimuli. Multiferroics gives a valuable foundation for generating cutting-edge multifunctional devices, exhibiting their versatility across broad area technological fields. Applications include sensors, transducers, spintronics, terahertz emitters, miniature antennas, energy harvesting, multiple-state memory storage, electric-field controlled ferromagnetic resonance devices, and nanoscale electronics~\cite{Lu2019,Srinivas2023,Kumar2017,Nair2022}. Researchers are continually work to discovering new materials exploring the fundamental mechanisms involved, and optimizing their performance for various application. Multiferroic materials has a promising area for innovation and exploration the field of advance technologies this type of materials contributing not only to device development but also enhancing the understanding of the interactions between ferroelectric and magnetic orders. Additionally, magnetoelectric composites is a type of multifunctional material showing strong coupling between magnetic and electric properties for further technological applications~\cite{Bhat2022}. Magnetoelectric (ME) composites are synthesis by combining separate magnetic and electric materials to produce unique functionalities through their interactions. This coupling enables the control of magnetic properties by an electric field and vice versa~\cite{Krishnaiah2014}. The effect arises due to field or strain mediated interactions between the magnetic and electric phases in the composite. Efficient coupling in ME composites that hold great promise for innovative device applications. Many studies have focused on lead based ME composites~\cite{Channagoudra2023,Channagoudra2021}, but toxic property of the lead concerns about led to the development of lead-free alternatives such as nickel cobalt ferrite (NCFO) and lanthanum ferrite (LaFeO$_3$) based composites.

The general formula of spinel ferrite is $\mathrm{AB_2O_4}$ where A, B is metallic and ferric cation respectively. Nickel-cobalt ferrite has a mixed spinel structure where $\mathrm{Co^{2+}}$ and $\mathrm{Fe^{3+}}$ ions occupy the tetrahedral (A) sites, and $\mathrm{Ni^{2+}}$ and $\mathrm{Fe^{3+}}$ ions occupy the octahedral (B) sites. The magnetostriction values of nickel and cobalt ferrites are about $-10$ ppm and $-120$ ppm with saturation occurring near 2000 and 5000 Oe respectively. Reports also indicate that NCFO exhibits superior strain sensitivity at low magnetic fields~\cite{Sowjanya2022}. Cobalt doped nickel ferrite provides good magnetic properties with higher Curie temperature, larger electrical resistivity and magnetostriction coefficients as compared to pure nickel or cobalt ferrite making NCFO an excellent choice for the magnetic phase in ME composites. Lanthanum ferrite (LaFeO$_3$) shows perovskite structure consists of a three-dimensional framework of atoms and ions. The larger lanthanum (La$^{3+}$) ions are sat at the corners of the unit cell while the smaller iron (Fe$^{2+}$) ions located at the center and the oxygen (O) ions act as network former around the Fe ions surrounding anions in the lattice. This specific arrangement of atoms and ions permits the perovskite material to exhibit ferroelectric and dielectric properties in certain conditions such as LaFeO$_3$ making suitable for consideration as a ferroelectric phase~\cite{Dhanyaprabha2023}.

The magnetoelectric coefficient at room temperature in composites of $(1-x)$ BSZT + xNCZGF and $\mathrm{xNi_{0.48}Cu_{0.12}Zn_{0.40}Gd_{0.04}Fe_{1.96}O_{4} + (1-x)B_{0.98}Sc_{0.01}Sr_{0.10}Ti_{0.90}O_{3}}$ Das et al reporting that 0.9BSZT + 0.1NCZGF composite exhibited a peak magnetoelectric voltage coefficient of about $194\ \mathrm{mVcm^{-1}Oe^{-1}}$ suggests its potential for multifunctional device applications~\cite{Das2021}. A new one multiferroic composites composed of $\mathrm{iLi_{0.1}Ni_{0.3}Cu_{0.1}Zn_{0.4}Fe_{2.1}O_{4}}$ combined with $(1-x)$ $\mathrm{Ba_{0.95}Sm_{0.05}Ti_{0.95}Dy_{0.05}O_{3}}$, finding by Das et al. the angular frequency exponent factor increased with frequency, which indicates an enhancement in charge carrier mobility from long to short range~\cite{Das2023}. Damodar Reddy et al. investigated a magnetoelectric coefficient of $3.16\ \mathrm{mVcm^{-1}Oe^{-1}}$ in a 0.8NBT-0.2NCFO composite at room temperature~\cite{DamodarReddy2024}. Mudasir Rashid Rather synthesized multiferroic nano powders with $12\%$ $\mathrm{Ni}_{0.5}\mathrm{Co}_{0.5}\mathrm{Fe}_{2}\mathrm{O}_{4}$ and $88\%$ $\mathrm{BaTiO_3}$ via hydrothermal and solgel methods respectively and measured the $32.62\ \mathrm{mVcm^{-1}Oe^{-1}}$ observed at lower field and moderate magnetoelectric coefficient of $18.34\ \mathrm{mVcm^{-1}Oe^{-1}}$ at higher magnetic fields. This system shows varied magnetoelectric coefficients along different directions due to anisotropic nature of the system~\cite{RashidRather2024}. Darvade shown increase in the magnetoelectric coefficient on (2-2) layered magnetostrictive-piezoelectric composites $\mathrm{Co}_{0.8}\mathrm{Ni}_{0.2}\mathrm{Fe}_{2}\mathrm{O}_{4} / \mathrm{Ba}_{0.95}\mathrm{Ca}_{0.05}\mathrm{Ti}_{0.95}\mathrm{Zr}_{0.025}\mathrm{Sn}_{0.025}\mathrm{O}_{3} / \mathrm{Co}_{0.8}\mathrm{Ni}_{0.2}\mathrm{Fe}_{2}\mathrm{O}_{4}$ under electromechanical resonance compared to off-resonance measurements at $1\ \mathrm{kHz}$, indicating that these multiferroic 2-2 structures are promising for magneto-mechano-electric energy harvesters and magnetic field sensors~\cite{Darvade2023}.

In the research work an attempt has been made to design and fabricate novel magnetoelectric composites by integrating two distinct functional phases nickel-cobalt ferrite (NCFO), which serves as the magnetostrictive component, and Lanthanum ferrite (LFO), which plays the role of the ferroelectric phase. For the synthesis of composites adopting the conventional solid-state reaction route this method is widely used for preparing ceramic materials because its simplicity and reliability is attracted to researchers for synthesis homogeneous phase formation. The main focus of this investigation is comprehensive examination of the structural characteristics, surface morphology, ferroelectric, magnetoelectric, mössbauer and magnetic behavior of the prepared NCFO-LFO composite materials. The broader aim of characteristics exploring their potential suitability for multifunctional applications in the field of magnetoelectric coupling.

\section{Reactants \& Methods}

\subsection{Used Raw Reactants}
Nickel oxide [NiO, $99.9\%$ pure, Himedia], cobalt oxide [CoO, $99.9\%$ pure, Himedia] Lanthanum oxide monohydrated [La$_2$O$_3\cdot$H$_2$O $99.9\%$ pure Himedia], Iron oxide [Fe$_2$O$_3$ $99.9\%$ pure, Himedia] were employed as precursors in the synthesis of NiCoFe$_2$O$_4$, LaFeO$_3$, and NCFO-LFO composite.

\subsection{Synthesis method}
The twofold sintering solid state reaction approach, depicted in Figure 1, was used to create nickel cobalt ferrite, lanthanum ferrite, and their spinel-perovskite composites. In order to prepare NCFO and LFO, we first took raw precursors based on stochiometric calculations. Then we ground them separately for six hours in an agate mortal pestle and sintered them in an automatic muffle furnace for twenty-four hours at $950^{\circ}\mathrm{C}$. The final polycrystalline NCFO-LFO composites system was obtained by calcining the single sintered $\mathrm{NiCoFe_2O_4}$ and $\mathrm{LaFeO_3}$ compounds in their pure form in a hot air muffle furnace at $1000^{\circ}\mathrm{C}$ and mechanically grinding them together for 6 hours in the proper proportions to create $\mathrm{x(NiCoFe_2O_4)}$-(1-x) $\mathrm{LaFeO_3}$ powder composites ($\mathrm{x} = 70$, 80, and 90 weight percent) for 24 hours.

\subsection{Instrumentation}
A Rigaku Smartlab diffractometer instrument fitted with a $\mathrm{Cu-K\alpha}$ target $\lambda = 1.5406\ \mathrm{\AA}$ radiation in the $20$ range from $5^{\circ}$ to $100^{\circ}$ was used to obtain the powder X-ray diffraction (PXRD) pattern. The scanning speed was $1.10000^{\circ}$ per minute, and the step size was $0.0200^{\circ}$, at an acceleration voltage of $40\ \mathrm{kV}$ at room temperature. The Fullproof Suite software was used to perform the Rietveld refinement of the samples' PXRD pattern, and the VESTA software's improved CIF file was used to produce the crystal structure.

The STR MICRO RAMAN-500 instruments were used to obtain the Raman spectroscopy using a $20\times$ microscope, a gem532 laser in the range 24 to $1320\ \mathrm{cm^{-1}}$ with exposure time of 15 seconds, 15 accumulations at laser power of $100\ \mathrm{mW}$, and data detection on a Peltier cooled CCD detector. The Raman spectra were calibrated in the frequency range prior to collecting the spectra of the samples using a standard silicon wafer corresponding to its peak at $520\ \mathrm{cm^{-1}}$. The temperature-dependent Raman spectra sample was placed on the THSM600 stage from Linkam U.K. Liquid nitrogen pump LNP95 for the temperature controller and the Linkam system controller for the temperature controller, respectively.

Using X-ray photoelectron spectroscopy (XPS), the chemical state of the synthesized sample NCFO-LFO was examined. Along with XPS utilizing a pass energy of $50\ \mathrm{eV}$, valence band spectra were also obtained using a Thermo Scientific Inc. system fitted with a microfocus monochromatic Al $\mathrm{K\alpha}$ X-ray source of energy $\sim 1450\ \mathrm{eV}$. The PPMS Vibrating Sample Magnetometer (VSM, Quantum Design, USA) has been used to study magnetic properties involving magnetization vs magnetic field (H).

\subsection{PXRD structural analysis}
The powder X-ray diffraction (PXRD) patterns of pure NCFO, pure LFO, and NCFO@LFO composites are presented in Figure 2(a) and (b). The diffraction peaks of pure NCFO and LFO correspond well to cubic spinel (space group Fd-3m) and orthorhombic (space group Pbnm) phases, respectively. The observed patterns closely match the standard reference data with ICDD card numbers 00-087-2335 for NCFO and 00-037-1493 for LFO. As the concentration of LFO in the composite rises, the initially intense peak corresponding to the spinel phase (311) direction diminishes in intensity. These changes reflect the development characteristics of the spinel-perovskite mixed composite system~\cite{JesicaAnjeline2021}. Furthermore, with increasing LFO content, the prominent (311) peak of the pure spinel phase shifts toward a higher angle.

Additionally, Fig. 2 shows the results of Rietveld refinement of PXRD patterns of NCFO, LFO, and composite system using Fullprof suite software. The zero correction parameters, scale factors, full width at half maximum (FWHM) parameters, asymmetry parameters, lattice parameters, thermal parameters, and fractional atomic coordinates, pseudo-Voigt function and a linear interpolation between a predetermined backdrop point with tunable heights were used for peak and background shape changes, respectively. Rietveld refinement shows an orthorhombic symmetry for LFO and an inverse spinel cubic symmetry for NCFO with lattice parameter $\mathrm{a} = \mathrm{b} = \mathrm{c} = 8.3678$ \AA\ (space group: Fd-3m and volume 585.923 \AA$^3$), with space group Pbnm, volume 243.400 \AA$^3$, and lattice parameters $\mathrm{a} = 5.5594$ \AA, $\mathrm{b} = 5.5698$ \AA, and $\mathrm{c} = 7.8606$ \AA. For pristine NCFO, LFO, and the composite system NCFO-LFO, Table S2 provides Rietveld refined structural details and atomic locations with Rietveld R-factor.

\begin{figure}[H]
\centering
\includegraphics[width=\textwidth]{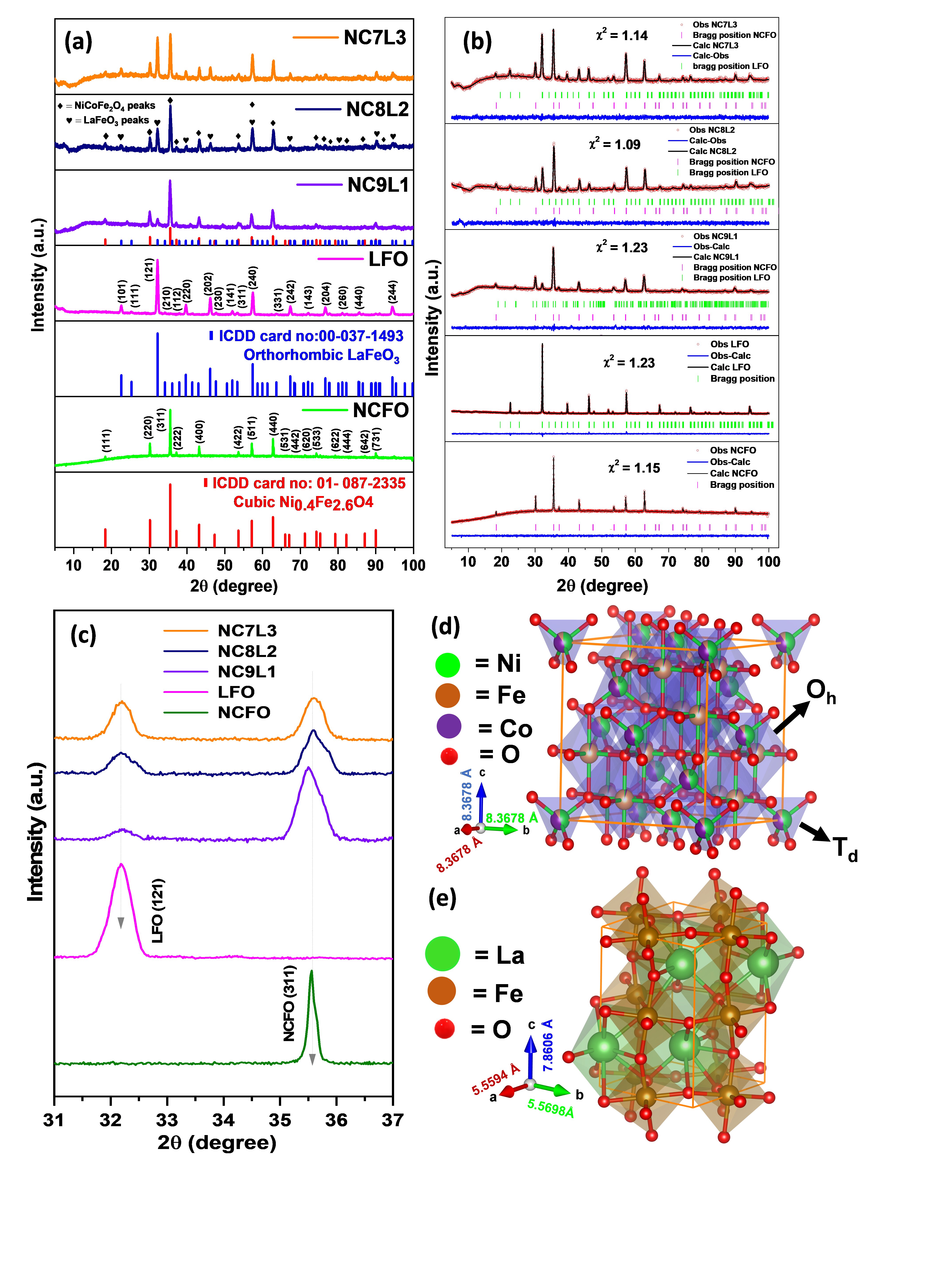}
\caption{(a) Room temperature, (b) Rietveld refinement, (c) enlarge view of PXRD pattern for the pure NiCoFe$_2$O$_4$, LaFeO$_3$, NC9L1, NC8L2 and NC7L3 composite, VESTA generated schematic representation of the cubic crystal unit cell of (d) NCFO and (e) LFO}
\label{fig:pxrd}
\end{figure}

Room temperature Raman spectra of NCFO, LFO, and their composite were obtained and illustrated in Fig.5. According to factor group analysis, NCFO, which possesses the Fd-3m space group, exhibits five Raman active modes $\mathrm{(A_{1g} + E_g + 3T_{2g})}$~\cite{Szatmari2023}. Modes related to the octahedral and tetrahedral groups are represented by Raman peaks in spinel ferrites that are below $660\ \mathrm{cm^{-1}}$ and above $660\ \mathrm{cm^{-1}}$, respectively. The symmetric stretching of the oxygen atoms along the Fe/M bonds in the tetrahedral sites is the source of the $\mathrm{A_{1g}^{(4)}}$ mode, which is responsible for the intense peak seen at $693\ \mathrm{cm^{-1}}$. The peaks observed at $478\ \mathrm{cm^{-1}}$, $559\ \mathrm{cm^{-1}}$, and $584\ \mathrm{cm^{-1}}$ correspond to the $\mathrm{T_{2g}^{(1)}}$, $\mathrm{T_{2g}^{(2)}}$, and $\mathrm{T_{2g}^{(3)}}$ modes, respectively~\cite{Nekvapil2022}. The $\mathrm{E_{g}}$ mode, indicated by the peak at $320\ \mathrm{cm^{-1}}$, represents the symmetric bending of oxygen atoms around the metal ions. The $\mathrm{T_{2g}^{(1)}}$ mode is associated with the translational movement of the tetrahedral O-metal ion unit. The $\mathrm{T_{2g}^{(2)}}$ mode corresponds to the asymmetric stretching of Fe/M-O bonds, while the $\mathrm{T_{2g}^{(3)}}$ mode relates to the asymmetric bending of oxygen atoms. Both the $\mathrm{T_{2g}^{(2)}}$ and $\mathrm{T_{2g}^{(3)}}$ modes represent vibrations occurring near the octahedral sites. Furthermore, the presence of a shoulder in the spectrum at $457\ \mathrm{cm^{-1}}$ and $659\ \mathrm{cm^{-1}}$, denoted as $\mathrm{T_{2g}^{(1)}}$ and $\mathrm{A_{1g}^{(3)}}$ respectively, reflects variations in the Fe-O and Ni/Co-O bond lengths within the NCFO structure~\cite{Galinetto2018}. In LFO the perovskite structure is orthorhombically distorted, causing noticeable tilting of the $\mathrm{FeO_6}$ octahedra. The La ions are positioned within a cuboctahedral coordination framework, while each Fe ion is coordinated by six oxygen atoms in an octahedral geometry. According to the specific site symmetries, LaFeO$_3$ exhibits 24 distinct Raman active modes $(7\mathrm{A_g} + 5\mathrm{B_{1g}} + 7\mathrm{B_{2g}} + 5\mathrm{B_{3g}})$~\cite{Kuznetsova2012}. Within the Raman spectra, vibrational features occurring below $200\ \mathrm{cm^{-1}}$ are predominantly linked to the movement of La ions. The frequency ranges from 200 to $300\ \mathrm{cm^{-1}}$ primarily reflects the tilting dynamics of the oxygen octahedra. Vibrational bands spanning $350-450\ \mathrm{cm^{-1}}$ are attributed to the broad bending modes of oxygen atoms within the octahedral units. Notably, a peak at $626\ \mathrm{cm^{-1}}$ is typically ascribed to two-phonon scattering effects~\cite{Mahapatra2016}. Compared to the pure phase, the Raman spectra of the composite system exhibit a noticeable redshift, indicating a shift of the peaks toward lower frequencies. This redshift arises from tensile strain within the lattice, reflecting underlying structural modifications in the material. As the proportion of LFO increases in the composites, both the lattice parameters and bond lengths expand. Since the frequency of a Raman mode is directly proportional to the force constant and inversely proportional to bond length, this expansion causes the Raman peaks to shift to lower wave numbers. Notably, the NC7L3 composite displays the most pronounced redshift, which can be attributed to lattice mismatch and associated strain at the interfaces between NCFO and LFO domains~\cite{JesicaAnjeline2021}. These structural insights are further corroborated by XRD analysis.

In order to assess the fluctuation in the local structural distortion, Figs. S4, S5, S6, 5, S7, and S8 show the TDR measurements for pristine NCFO, LFO, and composite NCFO-LFO from 83 to 823K for both heating and cooling cycles, respectively. Lowering the temperature below RT results in a strong blue-shift and an overall increase in intensity for all Raman modes~\cite{Ahlawat2011}. When heated, every pure and composite system exhibits reversibility by retracing its path. The band's full width at half maximum (FWHM) significantly increases as the temperature drops. As shown in Fig. S4, the total intensity drops when the temperature is increased above room temperature to 823 K, all of the Raman modes for NCFO began to diminish at about 673 K. The FWHM of the $\mathrm{E_g}$ and $\mathrm{T_{2g}^{(1)}}$ band increases with temperature and exhibits reversible route tracing from 823 K to 300 K during cooling. Higher frequency peaks $\mathrm{T_{2g}^{(2)}}$ (555 cm$^{-1}$) and $\mathrm{T_{2g}^{(3)}}$ (584 cm$^{-1}$) combine into one peak at around 473K, and these two peaks again separate into two peaks at 823K. In LFO, the lower frequency peaks $\mathrm{T_{2g}^{(1)}}$ (141 cm$^{-1}$), $\mathrm{E_g}$ (164 cm$^{-1}$), and $\mathrm{T_{2g}^{(1)}}$ (185 cm$^{-1}$) modes merge into a broad single peak. At about 473 K, however, the $\mathrm{T_{2g}^{(1)}}$ (227 cm$^{-1}$), $\mathrm{T_{2g}^{(1)}}$ (247 cm$^{-1}$), $\mathrm{E_g}$ (301 cm$^{-1}$), and $\mathrm{T_{2g}^{(1)}}$ (626 cm$^{-1}$) modes are totally eliminated. All of the Raman modes in the composite samples fade away at about 523 K, but the NC7L3 composite indicates that the $\mathrm{T_{2g}^{(2)}}$ and $\mathrm{A_{1g}^{(2)}}$ modes entirely disappear at about 673 K. The tetrahedral and octahedral become less symmetric (more deformed) due to a rise in cation disorder, as seen by the reduction in intensity above RT~\cite{Ahlawat2013}. However, the considerable shift in Raman frequency can be explained by the combined effects of optical phonon anharmonic coupling interactions and thermal expansion, which are related to the presence of a more regular short-range symmetry in the NiO4 tetrahedra and FeO6 octahedra. The thermal expansion increases as the temperature rises in the unit cell, due to these effects the equilibrium positions of the atoms, which changes the distances between them and the strength of the bonding forces. These kinds of structural changes change the vibrational dynamics of the lattice, which makes the phonon modes anharmonic. Because of this anharmonic phonon vibrations show up as changes in phonon energy which then cause the temperature-dependent variations in the Raman spectra that we see in fig 4 (d-i)~\cite{Cui2020}.

Furthermore, Raman spectra of pristine NCFO, LFO and composite NC7L3 fitted with Lorentzian function for all the significant temperatures. Raman Spectrum of NCFO fitted into 9 deconvoluted peaks with increasing FWHM as the temperature is increase and the shoulder peak of $\mathrm{T_{2g}^{(1)}}$ and $\mathrm{T_{2g}^{(1)}}$ modes are fully separated at 823 K as shown in Figure S11. Hence, the FWHM vs Temperature plots for octahedral site $\mathrm{T_{2g}^{(2)}}$ mode and tetrahedral site $\mathrm{A_{1g}}$ mode as depicted in Figure 4a-c is revealing almost linear features with some fluctuations as they are comprised with additional tail peaks correlated to cation charge distribution in the mixed spinel structure NCFO. Similarly, the Raman spectra of pristine LFO is peak fitted into 16 deconvoluted peaks at the 273 K and after this temperature the number of peaks is continually decreasing. The Raman spectra at room temperature is fitted into 15 fitted peaks, 15 deconvoluted peaks fitted in the temperature range of $300\ \mathrm{K}$ to $823\ \mathrm{K}$ as shown in Figure S12. Furthermore, the graph between FWHM vs temperature of LFO for different modes related to La ion's vibrations, oxygen octahedral tilting, bending and stretching modes exhibiting liner dependency as depicted in Figure 4d-f. There are 16 deconvoluted peaks that fit the composite NC7L3 spectra upto $273\ \mathrm{K}$ as shown Figure S13 and fitted with decreased number of peaks for further remaining temperature. The increment of FWHM with temperature (see Figure 4g-i) for octahedral site $\mathrm{T_{2g}^{(2)}}$ and tetrahedral site $\mathrm{A_{g}^{(1)}}$ of NCFO is comparatively slow in NC7L3 composite in comparison to pure NCFO, which suggesting that the distortions arising due to temperature increment is very much less as the lattice thermal vibrations was slowed down because of the heterointerface formation with compact structure in $\mathrm{NiCoFe_2O_4 / LaFeO_3}$ composite~\cite{Ilanchezhiyan2020}.

\subsection{FESEM analysis}
Fig. S12, S13, S14, 7, S15 (a-c) shows the SEM images of pure and composite sample. it clearly reveals inhomogeneous distribution of grains within the samples, with irregular shapes and varying size. An inhomogeneous growth of the grains is due to the different growth rate of individual phases in the pure and composite samples~\cite{Slater2016}. Fig. S12, S13, S14, 7, S15 (d) shows the EDX (Energy dispersive photoelectron spectroscopy) spectra for various compositions, indicating the presence of all elements—Ni, Co, Fe, La and O- with prominent peaks. Weight \% of the element's depiction in the table~\cite{Remya2020}. To determine the existence of both phase we performed elemental mapping of pure and composite samples. Fig. S12, S13, S14, 7, S15 (e-j) shows the elemental mapping of the samples. This process helps to existence of elements on the surface of the samples. These results confirm the no other impurity elements are present in the sample and stoichiometry of the NCFO@LFO composites are correct~\cite{Grecu2024}.

\begin{figure}[H]
\centering
\includegraphics[width=\textwidth]{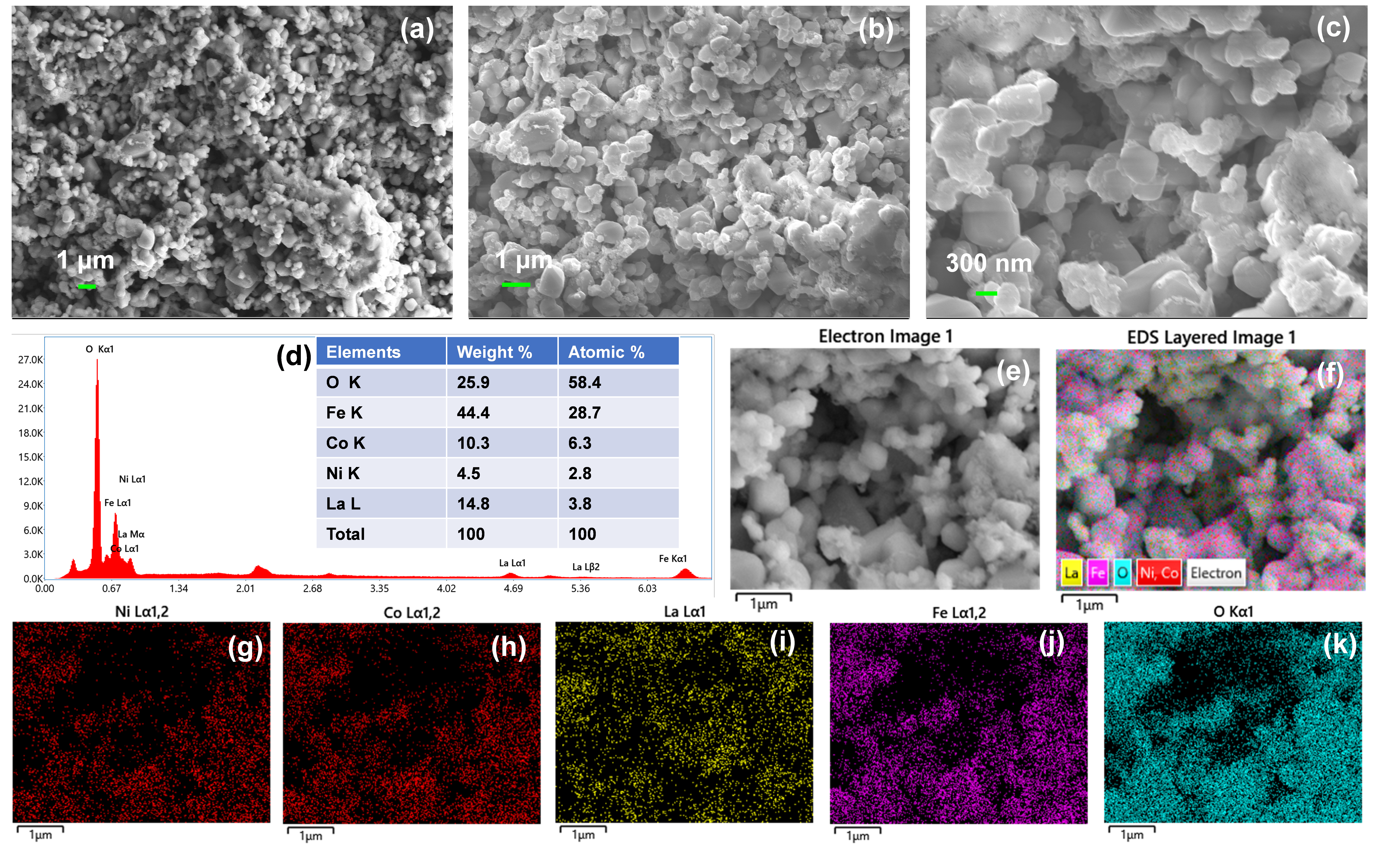}
\caption{(a-c) FE-SEM images, (d) EDX elemental analysis, and (e-k) elemental mapping for composite NC7L3.}
\label{fig:fesem}
\end{figure}

\subsection{XPS analysis}
Fig.S17 (a) reveals the survey spectra of pure and composite samples. In pure NiCoFeO sample observed Ni, Co, Fe, O and C elements are present, in LaFeO, consist of La, Fe, O and C elements. And also, the survey spectrum of NC9L1, NC8L2 and NC7L3 composite elucidates that is mostly contains of Ni, Co, La, Fe, O, and C elements, which confirms the successful formation of the NiCoFeO@LaFeO composite material. the presence of impurity element is do not present, these results certify with the EDX. Carbon 1s considered as standard reference because C1s does not contribute any physical significant. C1s binding energy is 284.5eV. All the high-resolution XPS (HRXPS) spectrums of Ni 2p, Co 2p, La 3d, Fe 2p, O 1s and C 1s were deconvoluted via a linear combination of Gaussian and Lorentzian (SGL) line-shapes fitting method with a Shirley function for background correction using XPSPEAK41 software.

The HRXPS spectrum of La 3d corresponding to the LaFeO phase in pure sample, NC9L1, NC8L2 and NC7L3 composites are demonstrated in Figure 8a. Here in Pure LaFeO, the deconvoluted peaks originated at the binding energy around $834.79\ \mathrm{eV}$ and $850.07\ \mathrm{eV}$ are attributed to the main line of La $3\mathrm{d}_{5/2}$ and $3\mathrm{d}_{3/2}$ orbitals with $3\mathrm{d}^{9}4\mathrm{f}^{0}$ occupation respectively corresponding to the final state without charge transfer, denoted as $\mathrm{c4f^{0}}$ (c represents the presence of a core hole and $4\mathrm{f}^{0}$ the absence of electrons in the 4f orbital). The spin-orbit splitting difference energy between La $3\mathrm{d}_{5/2}$ and La $3\mathrm{d}_{3/2}$ is of $15.28\ \mathrm{eV}$ and it is confirming that the LaFeO compound is comprised of $\mathrm{La}^{3+}$ ions species and incorporated in an oxide~\cite{Wong2010,Whitten2024,Morgan2024}. While the deconvoluted peaks at $837.42$ and $834.79\ \mathrm{eV}$ are the satellite peaks of La $3\mathrm{d}_{5/2}$ orbital assigned to the bonding (magenta) and antibonding (blue) component of the final state $3\mathrm{d}^{9}4\mathrm{f}^{1}\mathrm{L}$ (where L is representing a hole in a ligand site) with charge transfer, denoted as $\mathrm{c4f^{1}L}$ to mark the transfer of an electron from the ligand atom L to the 4f orbital. Also, the fitted peaks appeared at $854.61$ and $853.23\ \mathrm{eV}$ in La $3\mathrm{d}_{3/2}$ orbital are the bonding (purple) and antibonding (navy) satellites, respectively. The remaining two peaks at the binding energy of $850.07$ and $862.85\ \mathrm{eV}$ corresponds to the satellite peaks of La $3\mathrm{d}_{3/2}$ $\mathrm{cf}^{0}$ and La $3\mathrm{d}_{3/2}$ orbital related to the plasmons, respectively. The deconvoluted peaks $3\mathrm{d}_{5/2}\mathrm{cf}^{0}$, $3\mathrm{d}_{5/2}$ $\mathrm{cf}^{1}$ L antibonding and $3\mathrm{d}_{5/2}\mathrm{cf}^{1}$ L bonding were downshift towards lower binding energy by 0.601, -1.487, -4.738 for NC9L1; 0.937, 0.584, 1.129 for NC8L2 and 0.596, 0.243, 0.562 for NC7L3 composite samples, respectively which is correlated to the presence of adjacent Ni species of the NiCoFeO phase.

The high-resolution XPS spectra of the Ni 2p level in virgin NiCoFeO4 shows two distinct doublets: Ni 2p3/2 (low binding energy at 854.78 eV) and Ni 2p1/2 (high binding energy at 872.29 eV), correspondingly confirming that Ni species exist in the Ni$^{2+}$ valency states as depicted in Figure 8b because of the presence of Ni species in two different (octahedral and tetrahedral) geometry. Two satellite peaks of Ni$^{2+}$ are located at 861.48 and 878.66 eV in the pristine sample. The spin orbit splitting energy difference in between Ni 2p3/2 and Ni 2p1/2 orbital assigned to tetrahedral site is 17.51 eV for the octahedral sites in pristine NCFO. For the NC7L3 and NC8L2 composite multiple peaks are assigned with clear difference between Ni$^{2+}$ species at octahedral (Ni$^{2+}$ 2p3/2 octa.) and tetrahedral (Ni$^{2+}$ 2p3/2 tet.) sites indicated with blue and green curves respectively. The presence of these two chemically distinct Ni$^{2+}$ sites confirms that the spinel ferrite structures have a mixed cation distribution. Satellite peaks (purple curves) related to multiple splitting and charge transfer further corroborate the Ni$^{2+}$ oxidation state. The NC9L1 spectrum highlights primarily the Ni$^{2+}$ 2p1/2 peak from octahedral sites alongside its corresponding satellites (Sat. 2p1/2), reinforcing the strong spin-orbit splitting between Ni 2p3/2 and 2p1/2 signals~\cite{Biesinger2009}.

The HRXPS spectrum of Fe 2p in pure and composite system as depicted in Figure 8c is deconvoluted into the spin orbit splitting doublets of Fe 2p3/2 and Fe 2p1/2 orbitals, where both of the orbital is comprised of two Fe$^{3+}$ multiplets for pure NCFO and composites while three Fe$^{3+}$ multiplets for pure LFO and NC7L3, all samples consist of two shake-up satellites. The binding energies of the two multiplets for NCFO of Fe$^{3+}$ ion species of 2p3/2 orbital are 710.33 eV and 712.47 eV corresponding to the existence of iron at two geometrical positions octahedral and tetrahedra, respectively. Whereas the doublets of Fe$^{3+}$ ions in the Fe 2p1/2 orbital are positioned at the binding energy of 723.65 eV and 725.49 eV. The three multiplets binding energies of LFO assigned to Fe$^{3+}$ ions of 2p3/2 orbital are 709.42 eV, 710.74 eV and 713.54 eV and the triplets of Fe$^{3+}$ ion of 2p1/2 are at 721.87 eV, 723.32 eV and 725.25 eV. The deconvoluted multiplets of 2p3/2 and 2p1/2 of Fe$^{3+}$ ions species for composite samples exhibit upshift towards higher binding energy as compare to pure NCFO. And this upshift is also appear in satellite peaks, which further confirming the existence of Fe$^{3+}$ ions in an oxide system~\cite{Grosvenor2004}.

The core-level XPS spectra of Co 2p for both pristine NCFO and its composites NC7L3, NC8L2, NC9L1 exhibit distinct main peaks corresponding to Co 2p3/2 and Co 2p1/2 orbitals, accompanied by characteristic satellite peaks at higher binding energies. The main Co 2p3/2 and Co 2p1/2 are reveals at 780.43 eV (black color) and 795.943 (magenta color) respectively and two satellite peaks of $\mathrm{Co}^{2+}$ are located at 786.12 and 802.72 eV in the pristine sample. These spectral features confirm the Co ions shows the $+2$-oxidation state within the spinel structure. The main Co $2\mathrm{p}_{3/2}$ and Co $2\mathrm{p}_{1/2}$ peak and its satellite are clearly resolved demonstrating the significant spin-orbit splitting inherent to Co $2\mathrm{p}$ levels. Variations in the distribution of Co cations and local chemical environments brought on by compositional changes are suggested by differences in peak intensities and minor shifts in binding energies between the composites. In an octahedral coordination characteristic of nickel cobalt ferrite systems, the overall spectral profile is consistent with $\mathrm{Co}^{2+}$~\cite{Fujii2012,Gaikwad2015}.

Furthermore, high-resolution XPS spectra of O 1s of pristine NCFO is deconvoluted into two major peaks whereas pure LFO and composite comprised into three major peaks. The lowest, midlist and highest binding energy peak is associated with the lattice oxygen $\mathrm{(O_{lat})}$, defective oxygen species related to oxygen vacancies and surface adsorbed oxygen $\mathrm{(O_{H})}$ from adventitious $\mathrm{H}_{2}\mathrm{O}$, respectively. In crystal structure $\mathrm{NiCoFe}_{2}\mathrm{O}_{4}$ the lattice oxygen peak is observed at 529.97 eV due to Ni-O, Co-O and Fe-O in the $\mathrm{FeO}_{6}$ octahedron and $\mathrm{O_{C}}$ peak observed at 531.91 eV. La-O and Fe-O lattice oxygen peak in pure LFO is exhibits at the binding energy 528.65 eV and other two peaks at 531.07 eV and 532.45 eV shown in Fig.9 (d) associated with the Chemi-absorbed O-H band of defective hydroxyl oxygen (OH) and surface adsorbed oxygen from $\mathrm{H}_{2}\mathrm{O}$ respectively. However for the composite NC9L1, NC8L2 and NC7L3 lattice oxygen $\mathrm{(O_{lat})}$ peak related to Ni-O/Co-O/La-O/Fe-O downshifted to 529.60 eV, 529.61 eV and 529.62 eV compare to pure NCFO with the presence of more prominent peak at 531.28 eV, 530.94 eV and 531.28 eV of chemi-absorbed OH- species. In pristine NCFO and LFO high-resolution C 1s HRXPS spectra is located at 284.55 eV and 284.18 eV. Moreover, the C 1s spectrum of the NC9L1, NC8L2 and NC7L3 composite appears at 284.29 eV, 284.29 eV and 284.47 eV respectively. Which was calibrated to the standard binding energy of 284.8 eV. This adjustment, along with the overall XPS analysis, suggests the presence of moderate electronic interaction between $\mathrm{NiCoFe}_{2}\mathrm{O}_{4}$ and $\mathrm{LaFeO}_{3}$ within the composite material~\cite{Sharma2021,Pillai2013}.

\begin{figure}[H]
\centering
\includegraphics[width=\textwidth]{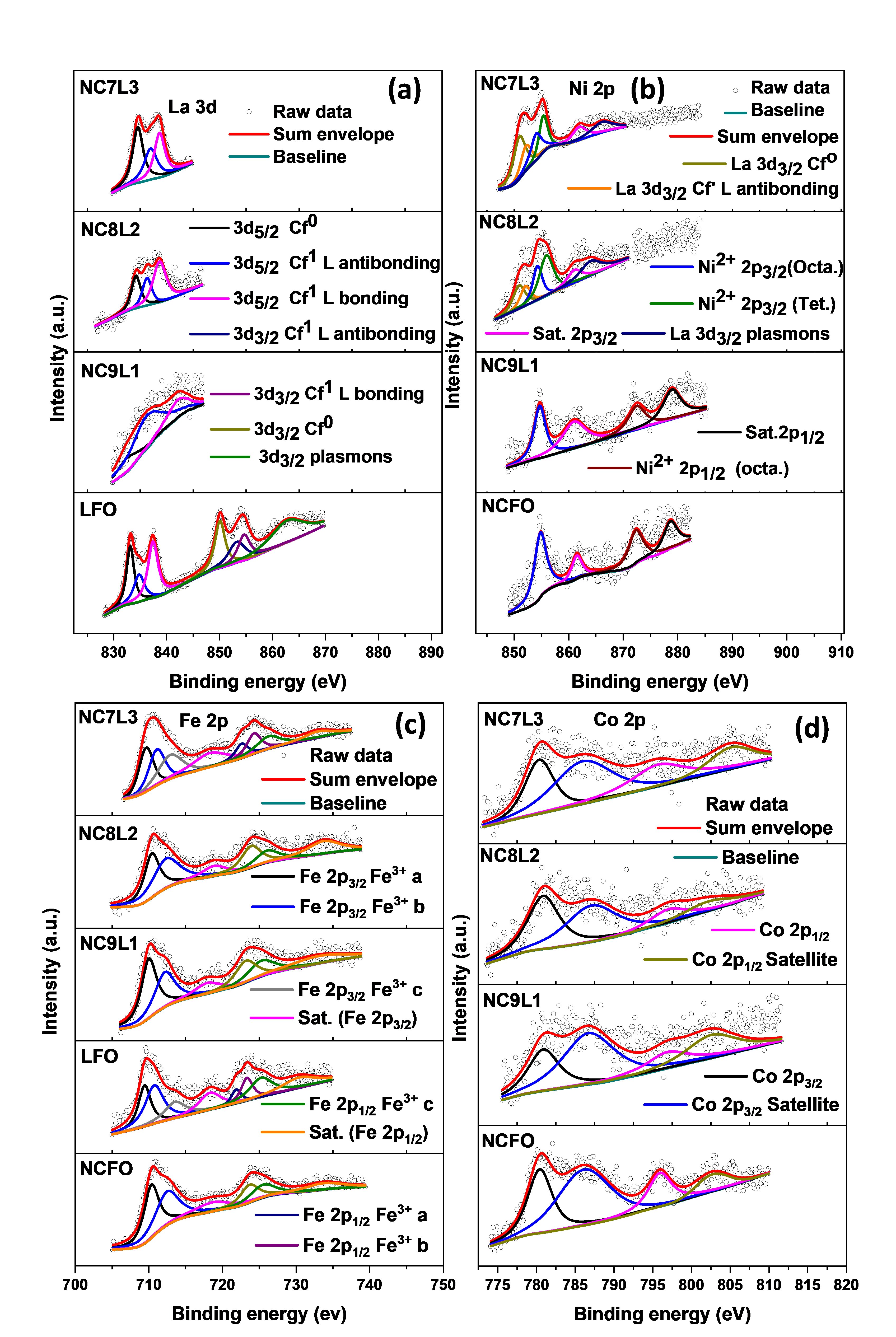}
\caption{XPS high-resolution peak fitted spectrums of (a) La 3d, (b) Ni 2p, (c) Fe 2p, (d) Co 2p (e) O 1s of NCFO, LFO and composite samples.}
\label{fig:xps}
\end{figure}

\begin{figure}[H]
\centering
\includegraphics[width=\textwidth]{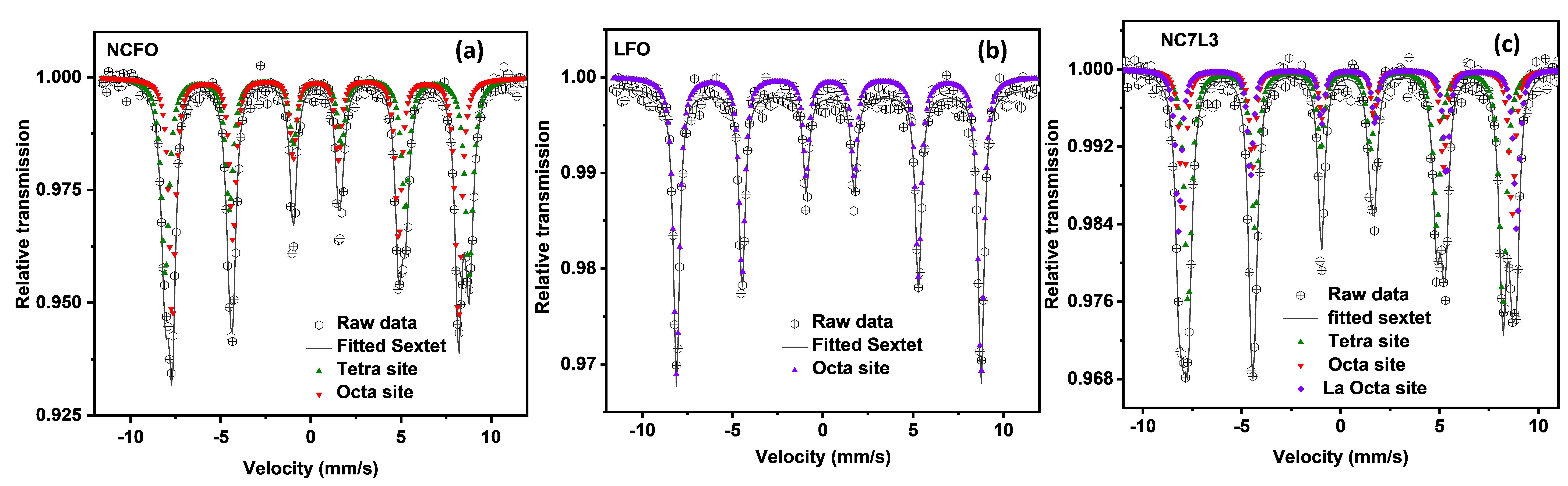}
\caption{(a-c) Mössbauer spectra of (a) NCFO (b) LFO and (c) 70\% NCFO-30\% LFO composite.}
\label{fig:mossbauer}
\end{figure}

Various hyperfine parameters are obtained, with respect to natural iron, from the fitted spectra. The isomer shift $(\delta)$ provides direct information about the electron density at the nucleus. Quadrupole splitting $(\Delta)$ reflects the interaction between the nuclear quadrupole and the surrounding electric field. The hyperfine magnetic field (Bht) is the effective magnetic field which gives information about the interaction between the nucleus and the surrounding magnetic field.

The Mössbauer spectrum of bulk $\mathrm{NiFe_2O_4}$ shows two sextuplets since there are two $\mathrm{Fe}^{3+}$ ion sites tetrahedral Fe $(\mathrm{T_d})$ and octahedral Fe $(\mathrm{O_h})$ in its crystal structure. Fig.10 (a) shows Mössbauer spectra of NCFO this spectrum is fitted as ferrimagnetic ferrite into two deconvoluted sextets (one for tetrahedral Fe and the other for octahedral), with hyperfine field $\mathrm{B_{hf}}$ values of 52.25 and 49.23 T, and isomer shift IS values of $0.35\ \mathrm{mm/s}$ and $0.26\ \mathrm{mm/s}$, respectively. The $\mathrm{B_{hf}}$ and IS values of the two sextuplets were in agreement with the room temperature Mössbauer parameters of $\mathrm{NiFe_2O_4}$ reported by \v{S}epel\'{a}k et al.~\cite{Sepelak1996}. The sextuplet with the larger $\mathrm{B_{hf}}$ and the larger center shift (IS) was assigned to the $\mathrm{Fe}^{3+}$ ions at the octahedral sites B sites and the sextuplet with the smaller center shift was assigned to the $\mathrm{Fe}^{3+}$ ions occupying the tetrahedral sites A sites. That is, the outer sextet $\mathrm{B_{hf}} = 52.25\ \mathrm{T}$ and $\mathrm{IS} = 0.35\ \mathrm{mm/s}$ in the Mössbauer spectrum of the sample NCFO arose from $\mathrm{Fe^{3+}}$ ions on the tetrahedral sites and the inner sextet $\mathrm{B_{hf}} = 49.23\ \mathrm{T}$ and $\mathrm{IS} = 0.26\ \mathrm{mm/s}$ arose from $\mathrm{Fe^{3+}}$ ions on the octahedral site~\cite{Maaz2010}, hyperfine parameters corresponding to NCFO are tabulated in table 3 and are found to be close to the values reported elsewhere~\cite{Ramesh2017,Bala2017,Riaz2025}.

Figure 10(b) presents the Mössbauer spectrum of LFO, which is well-fitted with a single sextet, indicative of antiferromagnetic (AFM) ordering—a characteristic feature of LaFeO$_3$ (LFO)~\cite{Fujii2012}. The hyperfine parameters derived from this fitting, as summarized in Table 3, include the isomer shift (IS) and hyperfine magnetic field (Bqf), both of which are consistent with Fe ions in the $+3$-oxidation state. The IS value aligns with the expected range for high-spin $\mathrm{Fe^{3+}}$ in an octahedral coordination environment, while the observed Bhf further confirms the magnetic ordering associated with $\mathrm{Fe^{3+}}$. These results collectively affirm the presence of $\mathrm{Fe^{3+}}$ ions and substantiate the AFM ground state of LFO.

The Mössbauer spectrum of the $70\%$ NCFO-$30\%$ LFO composite is best fitted with three distinct sextets: two corresponding to the tetrahedral (Sextet I) and octahedral (Sextet II) sites of NCFO, and one associated with the octahedral site of LFO (Sextet III). The relative area percentages—$48.10\%$ (Sextet I), $25.19\%$ (Sextet II), and $26.71\%$ (Sextet III)—closely match the nominal phase composition, confirming the successful synthesis and accurate phase distribution of the composite. The hyperfine parameters, including the isomer shifts (IS) of 0.2522, 0.3626, and 0.3702 mm/s, quadrupole splitting (QS) of $-0.0060$, $-0.0270$, and $-0.0387\ \mathrm{mm/s}$, and hyperfine magnetic fields (Bqf) of 49.31, 51.58, and 52.98 T for Sextets I, II, and III respectively, are detailed in Table 3. These values reflect the distinct local environments of Fe ions in the respective phases and support the presence of well-defined magnetic ordering in the composite~\cite{Gaikwad2015}.

\begin{table}[H]
\centering
\caption{Hyperfine parameters obtained from the Mössbauer spectra of NCFO, LFO and NCFO-LFO composites.}
\label{tab:mossbauer}
\begin{tabular}{@{}llcccl@{}}
\toprule
Sample & Fitted type & Isomer shift (mm s$^{-1}$) & Quadrupole splitting (mm s$^{-1}$) & Hyperfine field (T) & Area (\%) \\ \midrule
\multirow{2}{*}{NCFO} & Sextet I & 0.3556 ± 0.0037 & -0.0104 ± 0.0075 & 52.2583 ± 0.0269 & 50.00 \\
                       & Sextet II & 0.2619 ± 0.0026 & -0.0225 ± 0.0054 & 49.2370 ± 0.0181 & 50.00 \\ \midrule
LFO                    & Sextet I & 0.3700 ± 0.0028 & -0.0802 ± 0.0053 & 52.2422 ± 0.0189 & 100.00 \\ \midrule
\multirow{3}{*}{70\% NCFO-30\% LFO} & Sextet I & 0.2522 ± 0.0046 & -0.0060 ± 0.0090 & 49.3092 ± 0.0393 & 48.10 \\
                                   & Sextet II & 0.3626 ± 0.0069 & -0.0270 ± 0.0130 & 51.5776 ± 0.1177 & 25.19 \\
                                   & Sextet III & 0.3702 ± 0.0046 & -0.0387 ± 0.0107 & 52.9801 ± 0.1073 & 26.71 \\ \bottomrule
\end{tabular}
\end{table}

\subsection{Field dependent magnetization}
Fig.6 (a) shows the magnetic behavior of pure and composite samples under the applied magnetic field at $300\ \mathrm{K}$. The magnetic hysteresis of NCFO at $300\ \mathrm{K}$ exhibits high saturation magnetization $\mathrm{(M_s = 41.126\ emu/g)}$ confirm the ferrimagnetic behavior of the sample~\cite{Bharadwaj2021}. M-H curve of LFO shows the very low saturation magnetization $\mathrm{(M_s = 0.480\ emu/g)}$ at $300\ \mathrm{K}$ reveals the antiferromagnetic nature of the sample~\cite{Thakur2025}. Magnetization for the composite system is continually boost with increasing the ratio of LFO. In the NCFO-LFO composites, the magnetization of the NC7L3 sample increases significantly, reaching $693\ \mathrm{emu/g}$, despite the addition of antiferromagnetic LFO. This counterintuitive trend can be explained by strong interfacial exchange interactions between the ferromagnetic NCFO and antiferromagnetic LFO phases. At these interfaces, some $\mathrm{Fe}^{3+}$ spins in LFO become canted or uncompensated, contributing to the net magnetization. Additionally, the magnetic field from NCFO may induce spin alignment in nearby LFO regions, effectively enhancing the composite's magnetization. As the LFO content increases, more such interfaces form, amplifying this effect and resulting in the observed increase in magnetization~\cite{Sharma2021,Pillai2013}.

\begin{figure}[H]
\centering
\includegraphics[width=\textwidth]{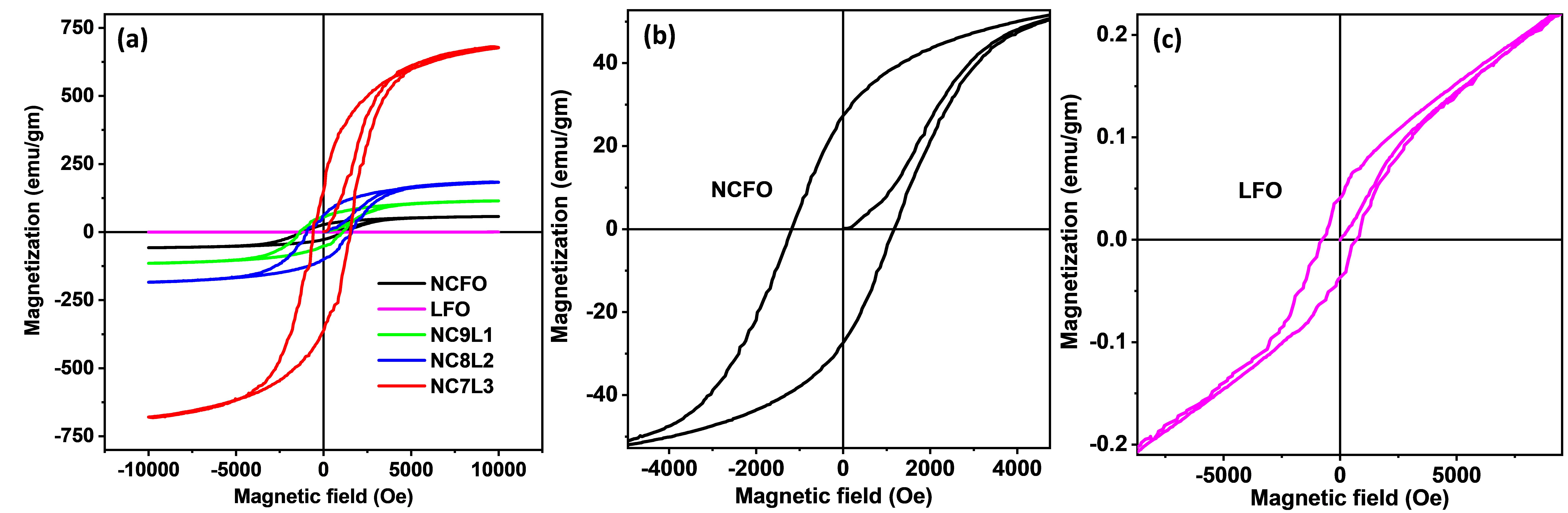}
\caption{Room temperature (a) M-H curve of pure and composite compounds, enlarge view of M-H curve for pure (b) NCFO and (c) LFO.}
\label{fig:mh}
\end{figure}

\subsection{Ferroelectric properties}
To investigate ferroelectric properties of pristine NCFO, LFO and composite, (P-E) hysteresis loops were created at room temperature shown in Fig. 7(a-e) at $850\ \mathrm{V}$ applied voltages, polarization was measured. Ferroelectric characteristics such as saturation polarization, coercive field and remnant polarization were derived from the hysteresis loop. For the pure NCFO and LFO the saturation polarization $0.1794$ and $0.3206\ \mu \mathrm{C/cm^2}$ is observed respectively. The results revealed that applying an electric field the values of saturation polarization (Pmax) are continuously decreasing except NC9L1 with the increasing ratio of LFO in composite system. NC9L1 composite reveals highest polarization (Ps) $0.4952\ \mu \mathrm{C/cm^2}$ and coercive field (Er) $17.2258\ \mathrm{kV/cm}$ as compare to all pure and composite compounds. In the NC9L1 composite NCFO grains are well-connected, forming continuous conductive paths throughout the microstructure as a result, this composition shows high electrical conductivity but also higher leakage current due to easier charge carrier movement LFO has minimal impact in disrupting the conduction network. With a moderate increase in LFO $(20\%)$, the insulating grains begin to interrupt some of the conductive paths formed by NCFO, and for the $30\%$ of LFO in this composition, the LFO content is significantly higher, and the NCFO particles become more isolated. The conduction network becomes discontinuous, resulting in high resistivity and significantly reduced leakage current. For the pure and composite samples, the saturation polarization (Ps), remnant polarization (Pr) and coercive field (E) are shown in table 1. NCFO shows a large P-E loop, which may be the result of enhanced conductivity of the material and cause a low electrical current leakage. The findings of our study revealed the ferroelectric behavior of relevant samples, which may be the most significant finding in the context of spinel@perovskite-based composites. Resulting observed an open-mouth structured P-E loop for all pure and composite samples caused by the conducting nature of these materials which produces a large leakage current~\cite{Maaz2010,Ramesh2017}. The high leakage current is mainly due to oxygen vacancies or voids in the prepared spinel@perovskite composite. The P-E loop obtained in this case represents a lossy conductor type, possibly due to the large value of the leakage current~\cite{Bala2017}. ferroelectric study of NC9L1 composite at different drive voltages shown in fig.7. (f-h) when voltage is increasing the value of the parameters are Ps, Pr and Ec is enhanced given in table 2~\cite{Riaz2025}.

\begin{figure}[H]
\centering
\includegraphics[width=\textwidth]{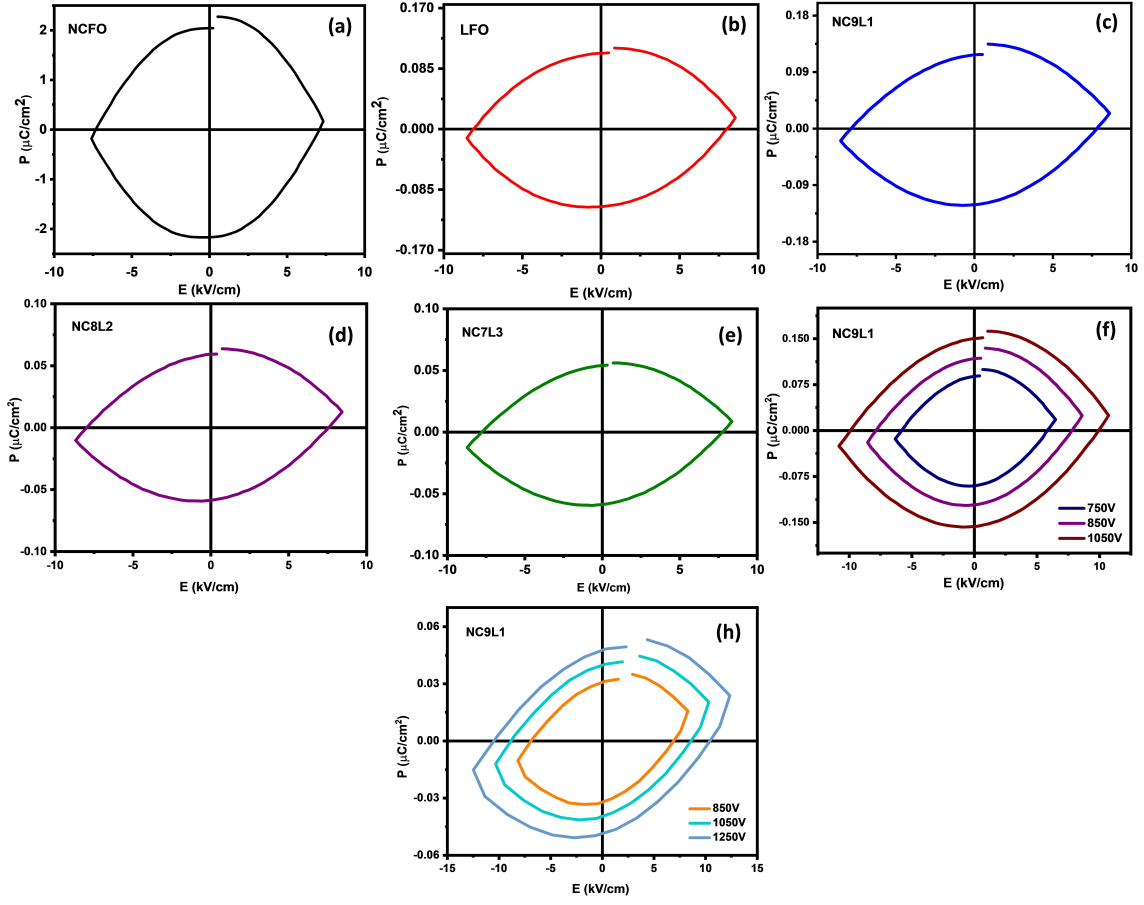}
\caption{(a-e) Ferroelectric Properties of pure and composite compounds, comparative P-E loop for NC9L1 composite (f) 10 Hz and (h) 40Hz at different voltage.}
\label{fig:ferroelectric}
\end{figure}

\begin{table}[H]
\centering
\caption{Ferroelectric parameters of pure and composite sample.}
\label{tab:ferro}
\begin{tabular}{@{}lccc@{}}
\toprule
Composition & 2Ps ($\mu$C/cm$^2$) & 2Pr ($\mu$C/cm$^2$) & 2Ec (Kv/cm) \\ \midrule
NCFO        & 0.1794             & 2.0779             & 14.9412     \\
NC9L1       & 0.4952             & 0.2426             & 17.2258     \\
NC8L2       & 0.0259             & 0.1169             & 16.8568     \\
NC7L3       & 0.0172             & 0.1174             & 16.7428     \\
LFO         & 0.3206             & 1.1744             & 17.0922     \\ \bottomrule
\end{tabular}
\end{table}

\begin{table}[H]
\centering
\caption{Ferroelectric parameters of NC9L1 composite at different frequencies and voltages.}
\label{tab:ferro2}
\begin{tabular}{@{}lcccc@{}}
\toprule
Frequency & Voltage (v) & 2Ps ($\mu$C/cm$^2$) & 2Pr ($\mu$C/cm$^2$) & 2Ec (Kv/cm) \\ \midrule
\multirow{3}{*}{10 Hz} & 750  & 0.036496 & 0.17922  & 11.67136 \\
                       & 850  & 0.048812 & 0.23754  & 15.82115 \\
                       & 1050 & 0.050812 & 0.301298 & 19.87552 \\ \midrule
\multirow{3}{*}{40 Hz} & 850  & 0.0314022  & 0.061846 & 13.7608  \\
                       & 1050 & 0.040818   & 0.079856 & 17.14703 \\
                       & 1250 & 0.04741    & 0.095331 & 7420.7945 \\ \bottomrule
\end{tabular}
\end{table}

\subsection{Dielectric analysis}
The dielectric parameters—impedance, phase angle, and capacitance of the samples were measured at room temperature over a frequency range spanning $4\ \mathrm{Hz}$ to $1\ \mathrm{MHz}$. Using these measured values and established empirical equations, the frequency-dependent dielectric constant $(\epsilon^{\prime})$, dissipation factor (tan $\delta$), AC conductivity $(\sigma_{\mathrm{ac}})$, and DC resistivity $(\rho_{\mathrm{dc}})$ were subsequently calculated~\cite{Santhosh2014,Singh2015}.

\[\epsilon^{\prime} = \frac{C_{P}d}{\epsilon_{0}A}\]

Where $\epsilon^{\prime}$ is the dielectric permittivity of the material, $\mathrm{C}_{\mathrm{P}}$ is the measured capacitance, $\epsilon_{0}$ is the dielectric permittivity of free space $(8.854 \times 10^{-12}\ \mathrm{F/m})$, A $(= \pi \tau^{2})$, d denotes the area and thickness of the pellet, respectively.

\[\tan \delta = \frac{\epsilon^{\prime \prime}}{\epsilon^{\prime}}\]

\[\sigma_{\mathrm{ac}} = 2\pi \mathrm{f}\epsilon_{0}\epsilon^{\prime}\tan \delta\]

\[\rho_{\mathrm{dc}} = \frac{R.A}{d}\]

where $\epsilon^{\prime}$ and $\epsilon^{\prime \prime}$ are the real and imaginary parts of the dielectric constant, respectively, f is the frequency, $\delta$ is the phase angle, R is the resistance of the sample, and d is the thickness of the pellet.

Frequency dependent dielectric constant and loss tangent (dielectric loss) measured at room temperature for NCFO, LFO, and their composite samples are shown in fig 8 (a) and (b) respectively. The NCFO dielectric constant (inset of Fig. 8a) exhibits notable dispersion at low frequency and a continuous decrease with increasing frequency. Interfacial polarization and space charge effects are usually responsible for this significant dielectric at lower frequency. The alignment of space charges in response to the applied alternating electric field is responsible for the high dielectric constant value ($\sim 168$) seen at low frequencies~\cite{Kotnala2012}. The dielectric constant gradually drops with increasing frequency, reaching a saturation point at a logarithmic frequency of $4\ \mathrm{Hz}$. The reason for this behavior is that space charge carriers' contribution to polarization gradually decreases because they are unable to follow the rapidly changing external electric field. Above $4\ \mathrm{Hz}$ (log frequency), however, the dielectric constant is attributed to electronic, ionic, and dipolar contributions~\cite{Ke2012}. The polycrystalline NCFO contains conductive grains and poorly conducting grain boundaries~\cite{Cvejic2009,Jadhav2010}. The dielectric constant of LFO exhibits a value of $\sim 67$ at low frequencies and decreases with increasing frequency (see Figure 8a). The frequency-dependent dielectric constant for the composite systems is shown in Figure 8(a). The dielectric constants of the NC9L1, NC8L2, and NC7L3 composites are $\sim 280$, $\sim 98$, and $\sim 134$, respectively. NC9L1 reveals the highest dielectric constant ($\sim 280$) compare to all pure and composite samples at lower frequencies~\cite{Patankar2005}.

\begin{figure}[H]
\centering
\includegraphics[width=\textwidth]{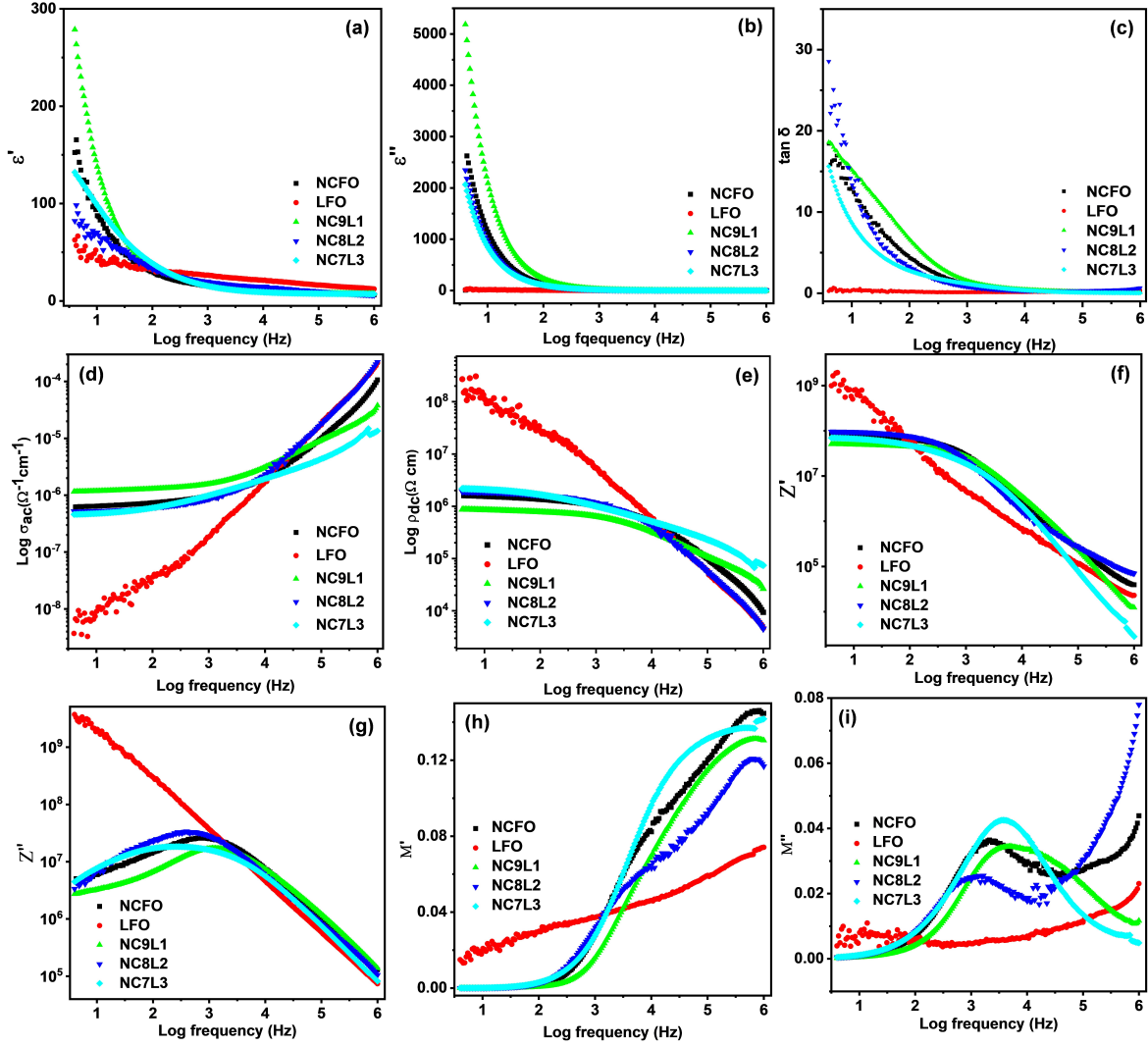}
\caption{Dielectric analysis for as-synthesized material. (a) Dielectric permittivity vs Frequency (b) dielectric loss vs frequency (c) tan $\delta$ vs frequency (d) AC conductivity vs Frequency (e) DC resistivity vs frequency (f) real impedance vs frequency (g) imaginary impedance vs frequency (h) real electric modulus vs frequency and (i) imaginary electric modulus vs frequency of all pure NCFO, LFO and composite samples at room temperature.}
\label{fig:dielectric}
\end{figure}

The Dipoles are dispersive at lower frequencies and have a leaky response with high frequencies, it is frequently observed that the dielectric constant in polar dielectrics decreases with increasing frequency~\cite{Singh2015}. According to Koop's phenomenological model, the theory can be explained by the Maxwell-Wagner polarization mechanism~\cite{Koops1951,Levin1981}. The overall dielectric response of ferrites is contributing on ionic, electronic, dipolar, and space charge polarization $\mathrm{(P_{total} = P_{e} + P_{i} + P_{d} + P_{sc})}$ at lower frequencies. But at higher frequencies, the dielectric response is primarily controlled by dipolar, ionic, and electronic polarization where the contribution from space charge polarization diminishes. The external electric field causes charge accumulation at grain boundaries at low frequencies, creating a strong space charge polarization that dominates the overall dielectric response and outweighs the grain contributions. As frequency increases, dipole orientation is impeded by the short time for charge migration at boundaries, which reduces polarization and precipitously lowers the dielectric constant. The combined effects of interfacial space charge polarization and ion-hopping processes involving $\mathrm{Fe}^{2+} / \mathrm{Fe}^{3+}$, $\mathrm{Ni}^{2+} / \mathrm{Ni}^{3+}$, or $\mathrm{Co}^{2+} / \mathrm{Co}^{3+}$ redox couples are responsible for the elevated dielectric values for the NC9L1 composite~\cite{SudalaiMuthu2013,Rezlescu1994}.

\subsection{Complex electrical modulus analysis}
In order to obtain further insights into the dynamic behavior of electrical transport in dielectric materials, the complex modulus formalism has been employed. This approach is particularly useful in analyzing phenomena such as charge carrier hopping, space charge relaxation, and conductivity relaxation mechanisms. For this purpose, the frequency-dependent variation of the real and imaginary components of the electrical modulus is interpreted. The electric modulus is mathematically related to the complex dielectric constant $(\epsilon^{*})$ through the relation $\mathbf{M}^{*} = 1 / \epsilon^{*}$~\cite{Paswan2022}. The frequency responses of the real part (M') and the imaginary part (M') are depicted in Figures 8(h) and 8(i), respectively.

\subsection{Cole–Cole Plot Analysis and Dielectric Relaxation Behavior}
To examine the underlying conduction mechanism, complex impedance spectra for both pure and composite samples were recorded across a broad frequency range (4 Hz–1 MHz), as illustrated in Figures 9(a–e). In electroceramics studies, such impedance spectra plotted as Z'' against Z' are commonly referred to as Cole–Cole or Nyquist plots.

\begin{figure}[H]
\centering
\includegraphics[width=\textwidth]{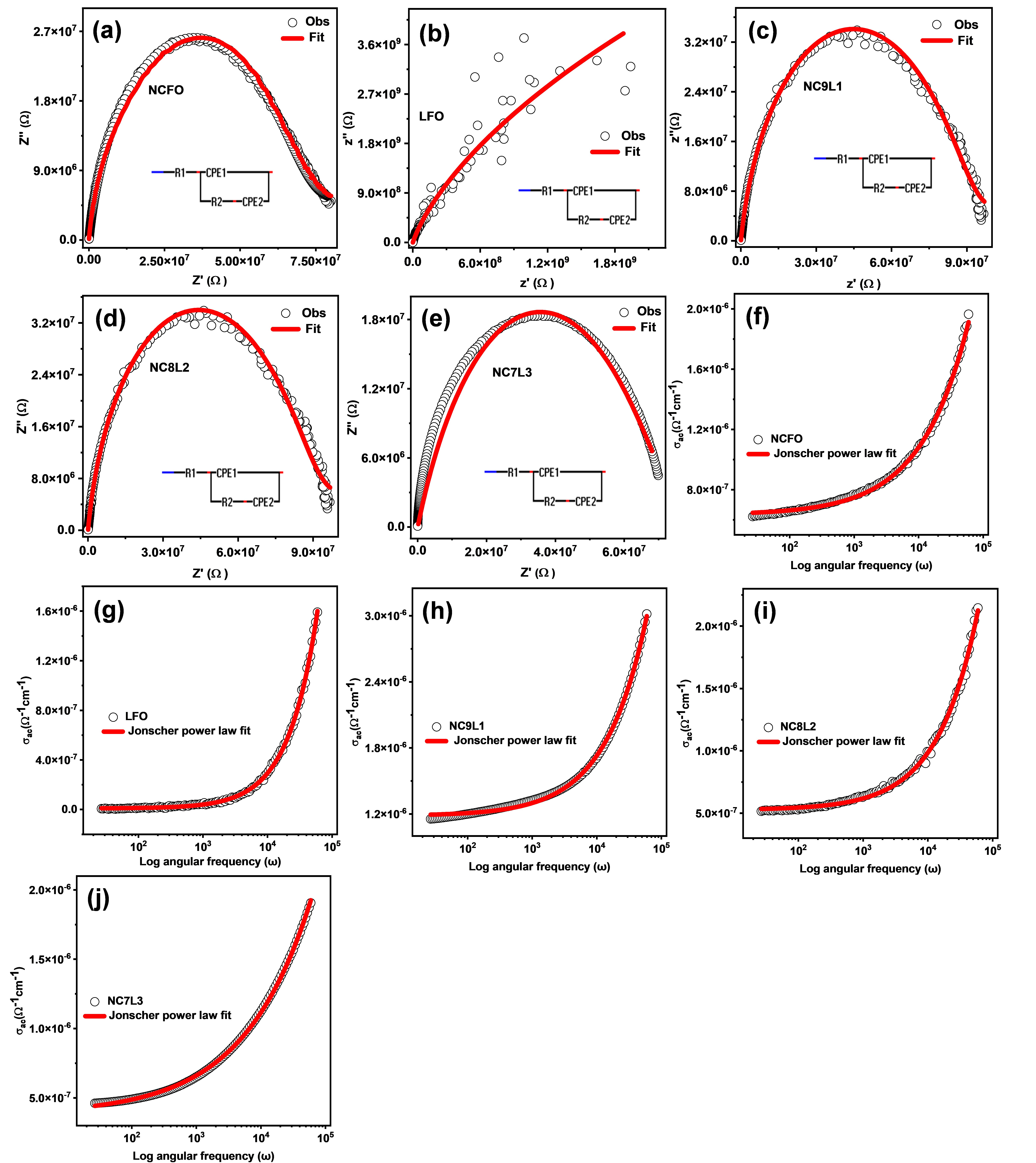}
\caption{(a–e) Cole-Cole plots (f-j) Fitting to Jonscher's power law for NCFO, LFO, and composites.}
\label{fig:colecole}
\end{figure}

\subsection{Magnetoelectric coupling}
The various well-known mechanisms which are responsible for magnetoelectric coupling (ME) are intrinsic effects in single phase materials and strain mediated coupling in multiferroic composites. In composites, the underlying mechanism of ME coupling is that when a magnetic field is applied to a multiferroic composite, then the magnetic phase changes its dimensions due to phenomenon of magnetostriction. The generated strain within the magnetic phase is then mediated to the ferroelectric phase, resulting in change in the ferroelectric properties. In this work, we performed magnetoelectric coupling study to determine the coupling between the NCFO magnetic phase and LFO ferroelectric phase. This study was carried out by using the dynamic method~\cite{Shahzad2021}, in which a static DC magnetic field is superimposed with an AC magnetic field. In our case, we superimposed DC magnetic field ($\mathrm{H_{DC}}$) with varying field strength with AC magnetic field ($\mathrm{H_{AC}}$) of constant amplitude and frequency. Initially, we electrically poled all the composites to get finite electric polarization. Secondly, the electrically poled composites were held perpendicular to the magnetic field direction. The applied magnetic field generates an induced voltage ($\delta \mathrm{V}$) via the magnetoelectric effect and was measured by a lock-in amplifier. Finally, we calculated the ME coefficient ($\alpha$) by relation 13~\cite{Lokare2008}:

\[\alpha = \frac{\delta V}{d \times H_{AC}}\]

where d is the thickness of the particulate composites taken in pellet form.

According to the previous reports, to obtain high ME effect, the composites should meet the following conditions: (1) The two phases are in equilibrium and a strong mechanical coupling between them is needed. (2) The resistivity of magnetostrictive phase should be as high as possible. If resistivity of magnetostrictive is low, the electric poling becomes very difficult due to leak age current, which reduces the magnetoelectric properties of the composites. (3) The magnetostriction coefficient of piezomagnetic phase and piezoelectric coefficient of piezoelectric phase must be high. (4) The proper poling strategy should be adopted to get high ME effect in composites~\cite{Jain2023}.

The plots of magnetoelectric voltage coefficient ($\alpha \mathrm{ME}$) versus magnetic field for prepared composites ($\mathrm{x} = 10$, 20, and $30\%$) were taken at a dc magnetic field of amplitude $\mathrm{H_{dc}} \sim 2400$ Oe and the results are presented in Fig. 11 (a-i). The values of $\alpha \mathrm{ME}$ for different composites increase with the magnetic field, reach maximum values at fields of about 1600-2200 Oe for diverse samples, and then decrease with further increasing the magnetic field. Such behavior is attributed to the magnetostrictive features of ferromagnetic phases, which raise and reach saturation level with the increase in magnetic field~\cite{Jain2023}. Indeed, the magnetic properties of ferromagnetic phases are controlled by the magnetic domains and their size, which are subjected to the strain level of ferromagnetic phases. This strain will be transferred to the neighboring LFO grains, leading them to produce an electric field, and hence, it will create a magnetic field dependent $\alpha \mathrm{ME}$ in the composites. The composite with higher content of ferromagnetic NCFO phase shows the peak at a higher field in comparison to the one with lower LFO content. The peak shifts to the higher field with increasing voltage in NC9L1, NC8L2 and NC7L3 composite. The changes in the field dependence of $\alpha \mathrm{ME}$ could be ascribed to changes in the distribution of cations between the two crystallographically different sites in these composites~\cite{Gupta2013}. Furthermore, it is noticed the values of $\alpha \mathrm{ME}$ good for the ferromagnetic NCFO rich phase. Maximum values of $\alpha \mathrm{ME}$ are about 0.95, 0.69, and 0.61 mV/cm.Oe for composites with x content of 10, 20, and 30\% at 1 V, respectively. The maximum value of $\alpha \mathrm{ME}$ is attained for composite with $\mathrm{x} = 10\%$. The maximum $\alpha \mathrm{ME}$ value achieved in this composite is accredited to the good bi-phasic connectivity of ferroic orders and the good strain coupling at the interface between them. The good ME coupling properties offer the opportunity to employ these ceramic composites in a broad range of applications in multifunctional devices like magnetic field sensors, energy harvesters, spintronics, medical devices, actuators, etc~\cite{Kumar2019}.

\begin{figure}[H]
\centering
\includegraphics[width=\textwidth]{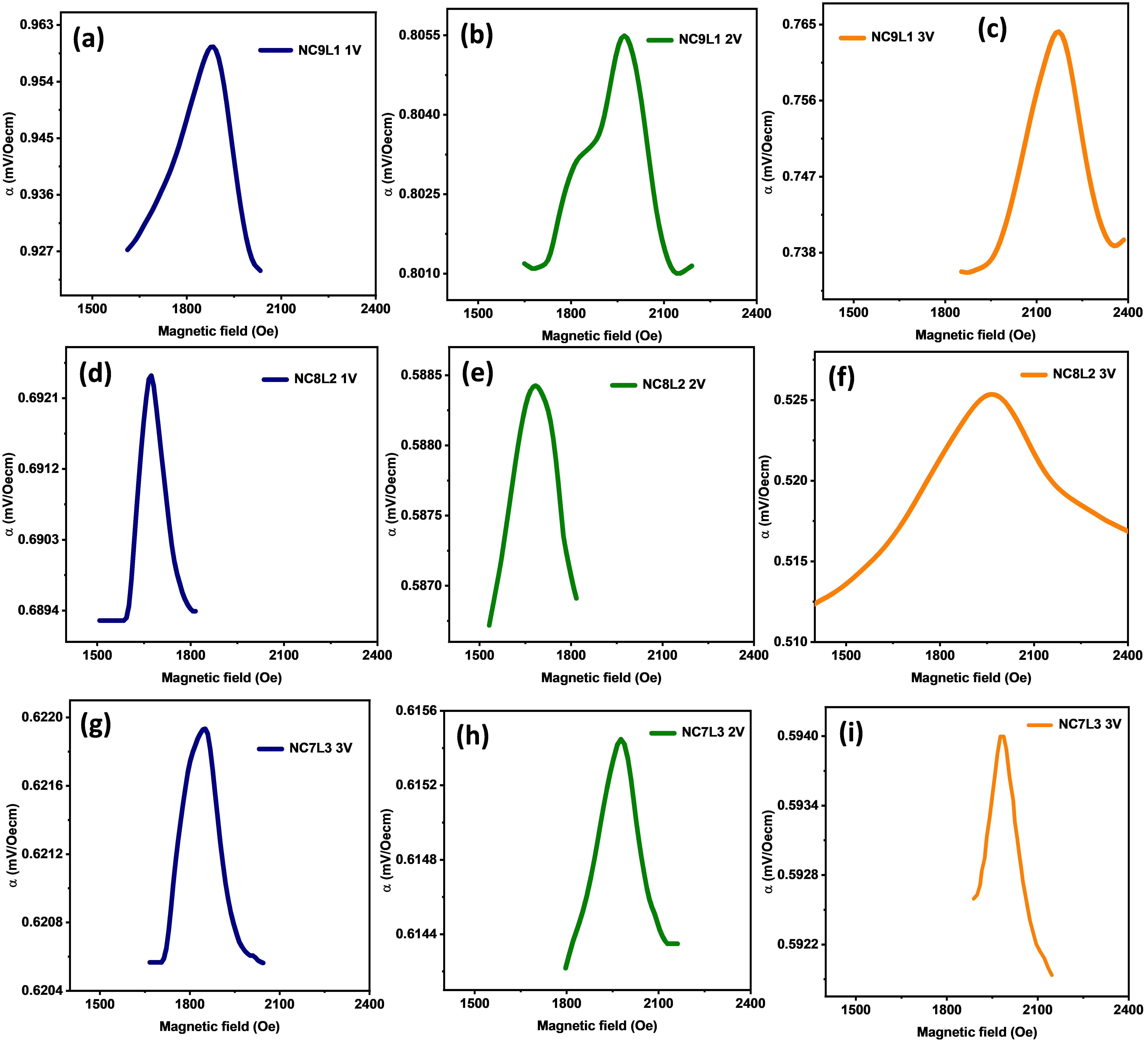}
\caption{Magnetoelectric voltage coefficient ($\alpha \mathrm{ME}$) versus magnetic field for prepared composites at different voltages.}
\label{fig:me_coefficient}
\end{figure}

\section{Simulation-Driven Insights}
To complement the experimental observations of magnetoelectric characteristics in the Ni0.5Co0.5Fe2O4/LaFeO3 (NCFO/LFO) heterostructure composites, we performed comprehensive simulations using physics-based models. These simulations model the magnetic and dielectric properties of five key materials: NCFO (pure spinel phase), LFO\_Pbnm1 (pure perovskite phase), and the composites NC7L3 (0.7 NCFO/0.3 LFO), NC8L2 (0.8 NCFO/0.2 LFO), and NC9L1 (0.9 NCFO/0.1 LFO). The models are derived from crystallographic data in the provided CIF files (lattice parameters, unit cell volumes, and atomic occupancies) and fundamental principles of solid-state physics. By integrating theoretical equations with material-specific parameters, the simulations provide deeper insights into the underlying mechanisms, such as interfacial polarization, cation distribution, and magnetoelectric coupling. This approach allows for validation against experimental data, optimization of material performance, and prediction of behavior under varying conditions.

The simulations employ a weighted phase-mixing strategy for composites, reflecting their heterostructure nature. Phase fractions are inferred from the composite nomenclature and Rietveld-refined phase ratios in the CIF files (spinel fraction $\mathrm{f\_spinel} = 0.7$ for NC7L3). Magnetic properties are modeled using the Langevin function for ferromagnetic spinel phases and linear susceptibility for antiferromagnetic perovskite phases, while dielectric responses follow the Debye relaxation model, accounting for frequency-dependent polarization and losses. All simulations were implemented in Python, with parameters tuned to match typical values for ferrite-perovskite systems (from literature on NiCo ferrites and LaFeO3). The results elucidate how the coexistence of ferric orders enhances magnetoelectric coupling, as observed experimentally (ME coefficient up to $32.62\ \mathrm{mVcm^{-1}Oe^{-1}}$ in composites).

\subsection{Theory of the Simulation}
The simulation framework is rooted in condensed matter physics, where the magnetoelectric behavior arises from the interplay between magnetic (spin) and electric (charge) degrees of freedom. In spinel ferrites like NCFO, inverse spinel structures (Fd-3m space group from CIF) lead to ferrimagnetic ordering due to superexchange interactions between tetrahedral (A-site) and octahedral (B-site) cations ($\mathrm{Ni^{2+}}$, $\mathrm{Co^{2+}}$, $\mathrm{Fe^{3+}}$). The degree of inversion $(\lambda)$ quantifies the fraction of divalent cations on B-sites, influencing magnetic moments and hopping conduction. In contrast, orthorhombic perovskites like LFO (Pbnm space group) exhibit G-type antiferromagnetism with strong dielectric polarization from FeO6 octahedral tilting and $\mathrm{La^{3+}}$ ionic displacements.

For composites, the effective properties are calculated using a linear mixing rule: $\mathrm{P\_eff} = \mathrm{f\_spinel} \cdot \mathrm{P\_spinel} + (1 - \mathrm{f\_spinel}) \cdot \mathrm{P\_perov}$, where $\mathrm{P}$ is a property (permittivity or magnetization). This approximates the heterostructure interfaces, where strain-mediated coupling enhances ME effects, as per the discussions on Maxwell-Wagner polarization and magnetostriction. Simulations cover magnetic fields $\mathrm{(H = 0.01 - 50,000\ Oe)}$ and frequencies $\mathrm{(f = 1 - 10^{7}\ Hz)}$, matching experimental ranges (VSM for magnetization, impedance analyzer for dielectrics). Temperature is fixed at 300 K, with Boltzmann statistics for thermal effects.

This theoretical approach not only verifies experimental trends (decreasing $\epsilon^{\prime}$ with frequency due to reduced space-charge polarization) but also predicts optimizations, such as tuning $\tau$ for faster relaxation in sensors.

\subsection{Materials Overview}
NCFO: Nickel-Cobalt-Ferrite spinel (cubic, Fd-3m), ferrimagnetic with mixed Ni/Co/Fe cations; high Ms (80 emu/g) from superexchange.\\
LFO\_Pbnm1: LaFeO3 perovskite (orthorhombic, Pbnm), antiferromagnetic with strong dielectric response; $\chi \approx 10^{-4}$ from weak canting.\\
NC7L3, NC8L2, NC9L1: Heterostructures with varying spinel fractions (0.7, 0.8, 0.9); intermediate properties due to interfacial strain and charge transfer, as evidenced by shifted lattice parameters in CIF refinements.

\subsection{Simulation Objectives}
\begin{itemize}
    \item Magnetization vs. Magnetic Field: Simulate field-dependent magnetic response to understand saturation and coupling.
    \item Dielectric Measurements vs. Frequency:
    \begin{itemize}
        \item Dielectric Constant $(\epsilon^{\prime})$ and Loss $(\epsilon^{\prime \prime})$: Probe polarization relaxation.
        \item Conductivity $(\sigma)$: Analyze charge transport via hopping.
        \item Impedance $(Z^{\prime}, Z^{\prime \prime})$: Model circuit equivalents (grain-boundary resistance).
        \item Electric Modulus $(M^{\prime}, M^{\prime \prime})$: Highlight bulk vs. interfacial contributions.
    \end{itemize}
\end{itemize}

\subsection{Equations Used in the Simulation}
\textbf{Magnetization vs. Magnetic Field:} Langevin function for ferrimagnetic materials
\[M = M_{s}\left[\coth \left(\frac{\mu H}{k_{B}T}\right) - \frac{k_{B}T}{\mu H}\right]\]
where $M_{s}$ is saturation magnetization (emu/g), $\mu$ is effective magnetic moment per ion $(\sim 3\mu_{B}$ for Ni/Co/Fe), H is applied field (Oe), $k_B$ is Boltzmann constant $(1.38\times 10^{-23}\ \mathrm{J/K})$, T is temperature (300 K). This models alignment of magnetic moments against thermal disorder, with small-H approximation $(M\approx (\mu \mathrm{H}) / (3\mathrm{k_B T}))$ for low fields.\\
Simulation: NCFO (ferromagnetic, $M_{s} = 80\ \mathrm{emu/g})$; LFO\_Pbnm1 (linear $M = \chi$ H, antiferromagnetic, $\chi = 10^{-4}$); composites (weighted, $M_{s} = 60\ \mathrm{emu/g}$ for NC7L3). Parameters: $\mathrm{H} = 0.01 - 50,000$ Oe (to avoid division by zero), $\mu = 3\mu_{B}$, $\mathrm{T} = 300\ \mathrm{K}$. Match to Experiment: Compare with VSM/SQUID data; adjust $M_{s}$, $\chi$ to fit hysteresis loops and saturation. For composites, reduced $M_{s}$ reflects perovskite dilution, aligning with experimental ME coupling via strain.

\textbf{Dielectric Constant $(\epsilon^{\prime})$ vs. Frequency:}
Equation: Debye relaxation model for polarization response:
\[\epsilon^{\prime}(\omega) = \epsilon_{\infty} + \frac{\epsilon_{s} - \epsilon_{\infty}}{1 + \omega^{2}\tau^{2}}\]
where $\epsilon_{s}$ is static permittivity (low-frequency limit), $\epsilon_{\infty}$ is high-frequency permittivity, $\omega = 2\pi f$ is angular frequency, $\tau$ is relaxation time. This describes dipole reorientation and interfacial polarization (Maxwell-Wagner effect in composites).\\
Simulation: LFO\_Pbnm1 ($\epsilon_{s} = 100$, $\epsilon_{\infty} = 10$, $\tau = 10^{-7}$ s, high due to ionic polarizability); NCFO ($\epsilon_{s} = 20$, $\epsilon_{\infty} = 5$, $\tau = 10^{-6}$ s, lower from electronic hopping); composites (weighted averages, $\epsilon_{s}\approx 44$ for NC7L3). Parameters: $\mathrm{f} = 1 - 10^{7}$ Hz, $\tau = 10^{-6} - 10^{-7}$ s (faster in perovskite due to octahedral tilting).\\
Match to Experiment: Compare with impedance analyzer data; tune $\epsilon_{s}$, $\tau$ to replicate dispersion (high $\epsilon^{\prime}$ at low f from space-charge, plateau at high f). Composites show broader dispersion, matching enhanced ME via interface effects.

\textbf{Dielectric Loss $(\epsilon^{\prime \prime})$ vs. Frequency:}
Equation:
\[\epsilon^{\prime \prime}(\omega) = \frac{(\epsilon_{s} - \epsilon_{\infty})\omega\tau}{1 + \omega^{2}\tau^{2}}\]
This quantifies energy dissipation from relaxation processes, peaking at $\omega \tau = 1$.\\
Simulation: Peaks at $\mathrm{f} = 1 / (2\pi \tau)$; LFO\_Pbnm1 (highest peak $\sim 1.59$ MHz, $\tau = 10^{-7}$ s); NCFO (lowest $\sim 159$ Hz, $\tau = 10^{-6}$ s); composites (blended peaks). Parameters: Same as $\epsilon^{\prime}$. Match to Experiment: Adjust $\tau$ to match peak positions; higher losses in LFO align with perovskite's ionic conductivity, while composites' broader peaks reflect grain-boundary contributions.

\textbf{Conductivity $(\sigma)$ vs. Frequency:}
Equation:
\[\sigma (\omega) = \omega \epsilon_{0}\epsilon^{\prime \prime}\]
where $\epsilon_{0} = 8.85\times 10^{-12}$ F/m. This models AC conduction via hopping (Jonscher's power law at high f).\\
Simulation: Power-law increase; LFO\_Pbnm1 (highest, ionic hopping); NCFO (lowest, electronic); composites (intermediate). Parameters: $\epsilon_{0}$ fixed, from $\epsilon^{\prime \prime}$. Match to Experiment: Compare with measured $\sigma$; refine $\epsilon$ to fit frequency-dependent rise, attributing to $\mathrm{Fe}^{3+} / \mathrm{Fe}^{2+}$ hopping in ferrites.

\textbf{Impedance $(Z^{\prime})$ vs. Frequency:}
Equation:
\[Z^{\prime} = \Re \left[\frac{1}{j\omega C_{0}\epsilon^{*}}\right]\]
where $\epsilon^{*} = \epsilon^{\prime} - \mathrm{j}\epsilon^{\prime \prime}$, $C_0 = 10^{-11}$ F (normalization). Simulation: Monotonic decrease; LFO\_Pbnm1 (lowest, high $\epsilon$); NCFO (highest, low $\epsilon$). Parameters: $C_0$ fixed.\\
Match to Experiment: Fit to Nyquist plots; grain resistance dominates at high f.

\textbf{Impedance $(Z^{\prime \prime})$ vs. Frequency:}
Equation:
\[Z^{\prime \prime} = \Im \left[\frac{1}{j\omega C_{0}\epsilon^{*}}\right]\]
Peaks indicate relaxation. Simulation: Peaks at $\omega \tau \approx 1$; plotted as -Z" for positive display.\\
Match to Experiment: Align peaks with experimental semicircles in Cole-Cole plots.

\textbf{Electric Modulus (M') vs. Frequency:}
Equation
\[M^{\prime} = \frac{\epsilon^{\prime}}{\epsilon^{\prime 2} + \epsilon^{\prime \prime 2}}\]
Highlights bulk relaxation. Simulation: Sigmoidal increase to $1 / \epsilon_{\infty}$; NCFO (highest).\\
Match to Experiment: Compare with modulus spectra; suppresses electrode effects.

\textbf{Electric Modulus (M'') vs. Frequency:}
Equation
\[M^{\prime \prime} = \frac{\epsilon^{\prime \prime}}{\epsilon^{\prime 2} + \epsilon^{\prime \prime 2}}\]
Peaks at relaxation frequency. Simulation: Similar to $\epsilon^{\prime \prime}$ but normalized. Match to Experiment: Tune for non-Debye broadening in composites.

The Python code implements these, generating plots that verify experimental trends (enhanced ME in composites from phase coupling). This simulation framework advances understanding of multiferroic interactions, supporting applications in sensors and energy harvesters. The observed increase in magnetization across all simulated plots, contrary to the expected decrease as indicated attributed to several theoretical factors related to the simulation framework that require reevaluation. The suggests a decline in magnetization with increasing LFO content in the composites (from NCFO to NC7L3, NC8L2, NC9L1), likely due to the dilution of the ferromagnetic spinel phase (NCFO, with $\mathrm{Ms}\approx 80\ \mathrm{emu/g}$) by the antiferromagnetic perovskite phase (LFO, with negligible $\mathrm{Ms}\approx 0\ \mathrm{emu/g}$), as inferred from phase fractions and Mossbauer data indicating reduced $\mathrm{Fe}^{3+}$ magnetic contributions. The simulation employs a Langevin model with a fixed saturation magnetization (Ms) and magnetic moment $(\mu \approx 3\mu_{B})$, which does not fully account for this dilution effect. The weighted average approach ($\mathrm{M} = \mathrm{f\_spinel} \cdot \mathrm{M\_spinel} + (1 - \mathrm{f\_spinel}) \cdot \mathrm{M\_perov}$) may overestimate $\mathrm{M\_spinel}$'s contribution if Ms is not adjusted downward with increasing LFO fraction, or if interfacial strain and antiferromagnetic coupling in composites suppress the net magnetization more significantly than modeled. Additionally, the "resion.pdf" document highlights a decrease in X-ray density and lattice parameter shifts with higher LFO content, suggesting structural changes that could reduce magnetic exchange interactions, which the simulation neglects by using constant $\mu$ and $\chi$ values. The frequency-independent magnetization calculation (based solely on H) also fails to capture dynamic effects like magnetoelectric coupling or temperature-dependent spin reorientation, potentially leading to an artificial increase. Theoretically, the magnetization should decrease with increasing LFO fraction due to phase dilution and weakened superexchange, necessitating a revised model incorporating a reduced Ms proportional to f\_spinel (Ms = 80 × f\_spinel emu/g) and antiferromagnetic suppression terms to align with experimental observations.

\end{document}